\documentclass[]{aastex631}

\newcommand{\sgra}{\mbox{SGR J1935}}
\newcommand{\sgrb}{\mbox{SGR J1935+2154~}}
\newcommand{\swift}{\mbox{\textit{Swift}}}
\newcommand{\swiftt}{\mbox{\textit{the Neil Gehrels Swift Observatory~}}}
\newcommand{\BAT}{\mbox{\textit{Burst Alert Telescope~}}}

\newcommand{\fermi}{\mbox{\textit{Fermi}/GBM}}
\newcommand{\gbm}{\mbox{\textit{GBM}}}
\newcommand{\chandra}{\mbox{\textit{Chandra~}}}
\newcommand{\XMM}{\mbox{\textit{XMM-Newton~}}}

\newcommand{\ep}{$E_{\rm p}$}

          \font\sixrm=cmr6

\def\teq#1{$\, #1\,$}                         
\def\rns{R_{\hbox{\sixrm NS}}}

\newcommand{\approptoinn}[2]{\mathrel{\vcenter{
  \offinterlineskip\halign{\hfil$##$\cr
    #1\propto\cr\noalign{\kern2pt}#1\sim\cr\noalign{\kern-2pt}}}}}

\shorttitle{Quasi-Periodic Spectral Oscillations from \sgra}
\shortauthors{Roberts, et al.}

\begin{document}

\title{Quasi-Periodic Peak Energy Oscillations in X-ray Bursts from \sgrb}

\correspondingauthor{Oliver Roberts}
\email{oroberts@usra.edu}

\author[0000-0002-7150-9061]{Oliver J. Roberts}
\affiliation{Science and Technology Institute, Universities Space and Research Association, 320 Sparkman Drive, Huntsville, AL 35805, USA.}

\author[0000-0003-4433-1365]{Matthew G. Baring}
\affiliation{Department of Physics and Astronomy - MS 108, Rice University, 6100 Main Street, Houston, Texas 77251-1892, USA}

\author[0000-0002-1169-7486]{Daniela Huppenkothen}
\affiliation{SRON Netherlands Institute for Space Research, Niels Bohrweg 4, 2333 CA Leiden, The Netherlands}

\author[0000-0002-5274-6790]{Ersin G\"o\u{g}\"u\c{s}}
\affiliation{Sabanc\i~University, Faculty of Engineering and Natural Sciences, \.Istanbul 34956 Turkey}

\author[0000-0002-1861-5703]{Yuki Kaneko}
\affiliation{Sabanc\i~University, Faculty of Engineering and Natural Sciences, \.Istanbul 34956 Turkey}

\author[0000-0003-1443-593X]{Chryssa Kouveliotou}
\affiliation{Department of Physics, The George Washington University, 725 21st Street NW, Washington, DC 20052, USA}

\author[0000-0002-0633-5325]{Lin Lin}
\affiliation{Department of Astronomy, Beijing Normal University, Beijing 100875, China}

\author[0000-0001-9149-6707]{Alexander J. van der Horst}
\affiliation{Department of Physics, The George Washington University, 725 21st Street NW, Washington, DC 20052, USA}

\author[0000-0002-7991-028X]{George Younes}
\affiliation{NASA Goddard Space Flight Center, University of Maryland, Baltimore County, Greenbelt, MD 20771, USA}

\begin{abstract}

Magnetars are young neutron stars powered by the strongest magnetic fields in the Universe (10$^{13-15}$~G). Their transient X-ray emission usually manifests as short (a few hundred milliseconds), bright, energetic ($\sim$~10$^{40-41}$~erg) X-ray bursts. Since its discovery in 2014, magnetar \sgrb has become one of the most prolific magnetars, exhibiting very active bursting episodes, and other fascinating events such as pulse timing anti-glitches and Fast Radio Bursts. Here, we present evidence for possible 42~Hz (24~ms) quasi-periodic oscillations in the $\nu F_{\nu}$ spectrum peak energy (\ep) identified in a unique burst detected with the Fermi Gamma-ray Burst Monitor in January 2022. While quasi-periodic oscillations have been previously reported in the intensity of magnetar burst lightcurves, quasi-periodic oscillations in the \ep~have not.  We also find an additional event from the same outburst that appears to exhibit similar character in \ep, albeit of lower statistical quality.  For these two exceptional transients, such \ep~oscillations can be explained by magnetospheric density and pressure perturbations.  For burst-emitting plasma consisting purely of $e^+e^-$ pairs, these acoustic modes propagate along a highly magnetized flux tube of length up to around $L\sim 130$ neutron star radii, with $L$ being lower if ions are present in the emission zone. Detailed time-resolved analyses of other magnetar bursts are encouraged to evaluate the rarity of these events and their underlying mechanisms. 

\end{abstract}


\section{Introduction} \label{sec:intro}

\sgrb (hereafter \sgra) was discovered in 2014~\citep{Stamatikos14} with the \BAT (BAT) on \swiftt (hereafter \swift), when the source emitted a few short X-ray bursts. Further observations of the source with the \swift~X-ray Telescope (XRT), \chandra and \XMM observatories, found a spin period of 3.24~s and a period derivative of 1.43$\times$10$^{-11}$ s$\cdot$s$^{-1}$, indicating a dipole magnetic field strength of $\sim$2.2 $\times$ 10$^{14}$ G~\citep{Israel16} at the equator, which confirmed the magnetar nature of the source.  The source has exhibited prolific bursting behavior since its discovery, with multiple burst-active episodes in 2015 and 2016~\citep{Younes2017}, 2018, 2019 and 2020~\citep{Lin2020}, and in 2021 and 2022~\citep{Roberts2023}. Comprehensive investigations of \sgra~bursts, using data from the $Fermi$ Gamma-ray Space Telescope's Gamma-ray Burst Monitor (\fermi) and \swift/BAT during these active episodes, showed that the magnetar became progressively more burst-active in every subsequent outburst episode~\citep{Lin2020}. The 2020 \sgra~outburst was the most energetic, producing a ``burst storm" on April 27$^{th}$ detected with multiple instruments, including \fermi~\citep{Kaneko2021} and NICER~\citep{Younes2020}, the latter reporting a burst rate maximum of 0.2 bursts/s over a $\sim$1200~s pointed observation. A hard burst with a $\nu$F$_{\nu}$ peak energy (\ep) reaching up to $\sim$80~keV on April 28$^{th}$, which was contemporaneous with a Fast Radio Burst \citep[FRB;][]{Ridnaia2021,CHIME2020,Bochenek2020,Mereghetti-2020-ApJ}, confirmed that at least one magnetar, \sgra, was a source of these exotic radio transients. \sgra~also underwent a radio-pulsar phase in October 2022, a few days after an anti-glitch (i.e., sudden spin down) was inferred from X-ray monitoring observations \citep{Younes2022}. Two additional FRB-like bursts from the source were detected during the same period~\citep{Kirsten2021}.

In this paper, we present \ep~oscillations found in two bursts from \sgra, detected on the 12$^{th}$ and 16$^{th}$ of January, 2022. These observations open up new territory for probing the physics of magnetar bursts and the character of neutron star magnetospheres. In Section~\ref{sec:data}, we introduce the data analysis methods and spectral analysis. In Section~\ref{sec:results}, we identify through visual inspection of the brightest bursts from \sgra~($f$ $\geq$ 1$\times$10$^{-6}$), two candidates that appear to exhibit \ep~oscillations. We present the time-integrated spectral fits for each burst, the time-resolved energy spectra, a periodicity search and temporal analysis. We present a theoretical interpretation of these results in Section~\ref{sec:discuss}, and summarize our findings in Section~\ref{sec:summary}.

\section{Data Analyses}
 \label{sec:data}

We use the triggered Time-Tagged Event (TTE) data (time resolution of $\sim$2 $\mu$s, 128 pseudo-logarithmically spaced energy channels), obtained with the \textit{Fermi} Gamma-ray Burst Monitor's (GBM's) thallium-doped Sodium Iodide (NaI) detectors with good detector-to-source angles ($\leq$60$^{\circ}$). These detectors have an effective spectral range of $\sim$8$-$900 keV, which nicely overlaps with the energy range of normal magnetar burst spectra ($\leq$300~keV). More information on the \fermi~instrument and its data types can be found in \citet{Meegan2009} and \citet{2020AvK}. 

\subsection{Spectral Analysis}

We performed spectral analysis on bright \sgra~bursts, emitted during the source activation starting in 2019 through 2022. While analyzing the \fermi~TTE data, we found two bursts (labeled throughout this study as ``Burst 1" and ``Burst 2"), that appear to show variations in their $\nu$F$_{\nu}$ peak or \ep~values. We performed a time-resolved spectral analysis of these bursts using a power-law with an exponential cutoff, commonly referred to as a Comptonized (COMPT) model. This model is defined as:
\begin{equation}
    F(E) = A\Biggl(\frac{E}{E_{\rm piv}}\Biggl)^{\alpha} \exp \left[-\frac{(\alpha+2)E}{E_{\rm p}}\right] ,
\end{equation}
where,
\emph{A} is the amplitude in ph/s/cm$^{2}$/keV,
\ep~is the $\nu$F$_{\nu}$ peak in keV,
\emph{$\alpha$} is the power-law index,
\emph{$E_{\rm piv}$} is the pivot energy in keV, fixed at 100~keV in this study.

In addition to performing time-integrated and time-resolved spectral analysis using the COMPT model, other models such as a simple power-law (PL), an Optically-Thin Thermal Bremsstrahlung (OTTB), a single black-body (BB) and a double black-body (BB+BB), were also used. These were fit to the burst spectra using RMFIT (v4.4.2), developed specifically for the analysis of \gbm~data\footnote{https://fermi.gsfc.nasa.gov/ssc/data/analysis/rmfit}. We selected background intervals devoid of bursts several seconds before and after each event, fitting each interval with a polynomial function to model the background during the burst, which was subsequently subtracted. The Detector Response Matrices (DRMs) for triggered bursts are available as a part of the publicly accessible data products. For untriggered bursts, we generated DRMs using GBMRSP v2.0. For the spectral analysis of all the bursts in our sample, we used the NaI-detector energy range of $8-900$\,keV. We neglect an energy interval of $\sim$4 keV centered around 35~keV, where the k-edge feature from Iodine in the NaI detectors appears in the data, as this has not been modeled perfectly~\citep{Bissaldi2009}. The exclusion of this small portion of data from our analysis does not affect the results, but improves the statistics. 

We use the Bayesian Information Criterion \citep[BIC;][]{Schwarz1978,Liddle2007} to determine preferred models for the bursts in this study, which is often used for model comparisons with the maximum likelihood statistics. We calculated BIC for each spectral fit as follows:
\begin{equation}
  {\rm BIC} = -2 \ln \mathcal{L}_{\rm max} + k \ln N = {\rm CSTAT} + k \ln N , 
\end{equation}
where $\mathcal{L}_{\rm max}$ is the maximum likelihood, $k$ is the number of free parameters in the spectral model and $N$ is the number of data points. We then calculated $\Delta$BIC for each pair of the four models for comparing the posterior probabilities of the given two models. We employ $|\Delta$BIC$| \geq12$ to select a preferred model, which corresponds to the Bayes factor of $\sim$400, indicating that the posterior probability of the model with a smaller BIC value is higher by $>99.7\%$ \citep{KassRaftery1995, anderson2004, Liddle2007}.

We only report the parameters from the COMPT and BB+BB model fits to our data, as they fit the data better than the PL, OTTB and BB models. We compare the BIC values between the COMPT and BB+BB model fits by defining $\Delta$BIC = BIC$_{\rm COMPT}$ -- BIC$_{\rm BB+BB}$, where a BB+BB model is the preferred fit to the burst spectrum when $\Delta$BIC $\geq$ 12, and a COMPT model is the preferred fit if $\Delta$BIC $\leq -12$, at 3-$\sigma$ significance. The duration and spectral parameters from fitting the candidates using the aforementioned models, are shown in Table~\ref{tab:1}.

\subsection{Localization and Burst History}

In the following sections we discuss the time-history and localization of the two burst outliers detected in January 2022. The most recent event (Burst 1; see Table~\ref{tab:1}) was localized to a Right Ascension (RA) and Declination ($\delta$) of 294.2$^{\circ}$ and 23.8$^{\circ}$ respectively, within the 1-$\sigma$ error (3.6$^{\circ}$) of the known position of \sgrb (RA: 293.7$^{\circ}$, $\delta$: 21.9$^{\circ}$~\citep{Israel16}). Similarly, the other event (Burst 2, which occurred around 3.8 days earlier than Burst 1) was localized to RA: 291.8$^{\circ}$, $\delta$: 24.5$^{\circ}$, with an error of 4.5$^{\circ}$ (1-$\sigma$), 3.3$^{\circ}$ degrees from the \sgrb position. Both events are consistent, within their errors, with an origin from \sgra; they occurred during a heightened period of burst-activity from the source, detected by multiple instruments (i.e.,~\citet{Roberts2022, Ridnaia2022, Kozyrev2022GCN}). We used a Bayesian block algorithm, similar to that used in~\citet{Lin2020}, to determine the duration of each burst ($\tau_{BB}$), shown in Table~\ref{tab:1}.
 
\section{Results}
 \label{sec:results}

The time-integrated COMPT and BB+BB spectral fits for Burst 1 and Burst 2 are presented in Table~\ref{tab:1}, noting that the spectral analysis for Burst 1 was performed separately for the whole event (which includes the faint, weak emission), as well as solely for the brighter, second emission episode. The fits were performed over the energy range \teq{10-1000}keV, though most burst counts above background were at energies \teq{\lesssim 100}keV.  We find that a COMPT model fit to the spectrum is preferred for both candidates.

\subsection{\sgra~\ep~Oscillation Candidates}
 \label{sec:oscill}

\begin{figure}
    \centering
    \includegraphics[width=0.49\textwidth]{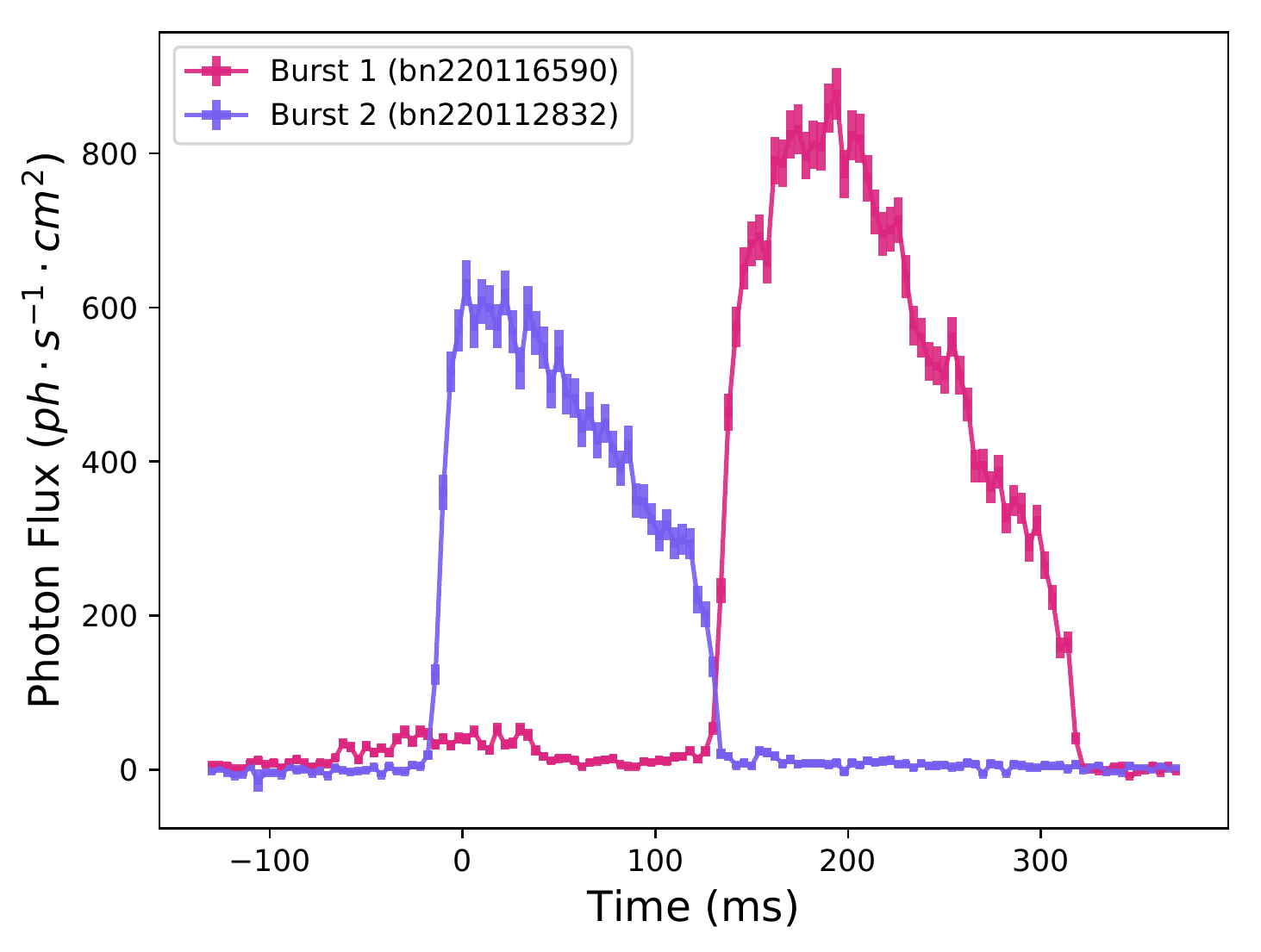}
    \includegraphics[width=0.49\textwidth]{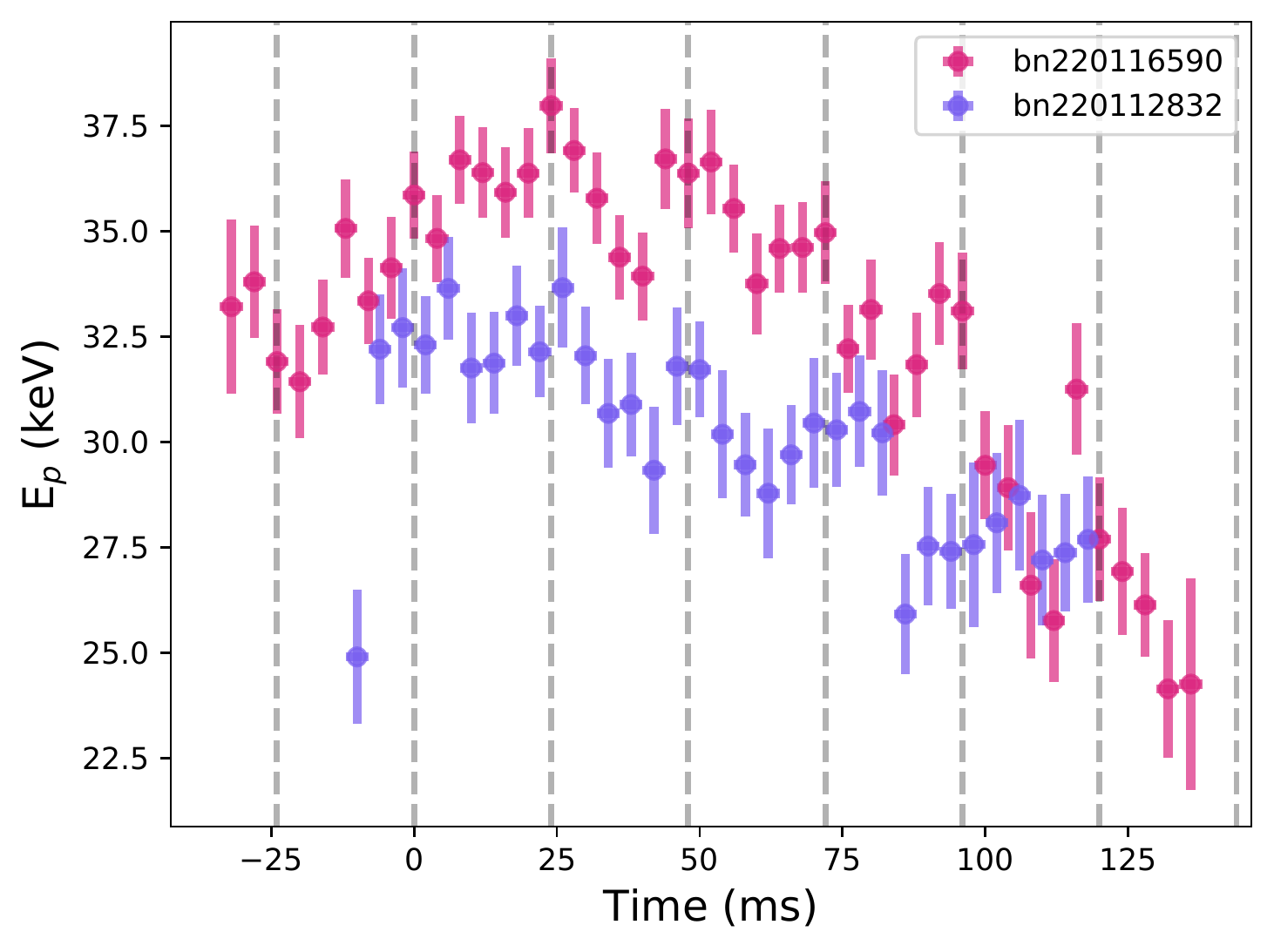}
    \caption{\textbf{Left:} The lightcurves of Burst 1 (bn220116590) and Burst 2 (bn22012832) plotted alongside each other within a 500~ms window starting at -130~ms. The trigger time for each event is the zero time. The first, weaker emission episode in Burst 1 lasts about 100~ms, starting at $\sim$125~ms. \textbf{Right:} A $\sim$175~ms window showing the \ep~behavior of Burst 1 (pink) and Burst 2 (purple). Burst 1 is shifted in time by -166~ms in order to show the similar \ep~oscillating frequency of 24~ms (42~Hz), highlighted by the dashed grey lines. The temporal binning for both panels is 4~ms.\label{fig:1}}
\end{figure}

The best example of fluctuations in \ep~is Burst 1, which triggered \fermi~at 14:09:38.94 UTC on January 16$^{th}$, 2022, and subsequently assigned the burst number, bn220116590. The burst appears in the data as a faint pulse followed about 100~ms later by a very bright pulse, shown in the left panel of Fig.~\ref{fig:1}, with a total duration of 693~ms. The time-integrated spectral results for both the whole burst and the brighter, second emission episode are shown separately in Table~\ref{tab:1}. We use NaI detectors 0, 1, 2, 3 and 5, as these all have good detector-source angles ($\leq$60$^{\circ}$). Upon fitting the brighter, second pulse, we find that the overall \ep~has a soft-hard-soft trend over that entire emission duration. When using finer temporal binning, the spectrum suggests possible periodic perturbations in \ep~(right panel of Fig.~\ref{fig:1}). Using the distance to the supernova remnant G57.2+0.8 (purported to be the host of \sgra), of 9~kpc~\citep{pav2013,Zhong2020}, we derive E$_{iso}$ and L$_{iso}$ values of 3.9$\times$10$^{40}$ erg and 1.9$\times$10$^{41}$ erg/s, respectively for the second, 200~ms emission episode. 

The next best example of an event with fluctuations in \ep~is Burst 2, which triggered \fermi~at 19:58:04.04 UTC on January 12$^{th}$, 2022, and was subsequently assigned the burst number, bn22012832. Burst 2 appears as a bright burst with a duration of 254~ms. We use NaI detectors 0, 1, 3 and 5, as these all had good detector-source angles ($<$60$^{\circ}$). When fitting a COMPT model to the spectrum, the general \ep~behavior of the burst is a quick rise followed by a slow decay, with weak perturbations on a similar timescale to that observed in Burst 1 (see Fig.~\ref{fig:1}). The time-integrated spectral results for this burst are shown in Table~\ref{tab:1}. Using the distance to \sgra~of 9~kpc, we derive E$_{iso}$ and L$_{iso}$ values of 2.2$\times$10$^{40}$ erg and 8.8$\times$10$^{40}$ erg/s, respectively for this 254~ms event. 

\begin{deluxetable}{cccccccccccc}[ht]
\tabletypesize{\footnotesize}
\tablecaption{\textbf{E$_{p}$ Oscillation Candidates}
}
\tablewidth{0.9\columnwidth}
\tablehead{
\colhead{ID} & \colhead{Date} & \colhead{Trigger Time} & $\tau_{BB}$ & \colhead{Fitted Interval (ms)}  & \multicolumn{2}{c}{Comptonized (COMPT)} & \multicolumn{2}{c}{BB+BB} & \colhead{$\Delta$BIC} & \colhead{Fluence} \\
& \colhead{YYMMDD} & \colhead{UTC}  & (ms) & \colhead{ t$_{start}$:t$_{stop}$} &\colhead{$\alpha$} & \colhead{\ep~(keV)} & \colhead{$kT_{1}$ (keV)} & \colhead{$kT_{2}$ (keV)} &  & \colhead{($\times$10$^{-6}$ erg\,cm$^{-2}$})}
\startdata
1 & 220116 (i) & 14:09:38.94 & 693 & $-$88 : 320  & 0.38$\pm$0.05 & 33.0$\pm$0.2 & 5.5$\pm$0.2 & 10.8$\pm$0.2 & $-$16  & 1.95$\pm$0.01 \\
& 220116 (ii)& - & - &  120 : 320 & 0.47$\pm$0.05 & 34.0$\pm$0.2 & 5.9$\pm$0.2 & 11.0$\pm$0.2 &  - & 3.98$\pm$0.03 \\
2 & 220112 & 19:58:04.04 & 254 & $-$24 : 276  & 0.51$\pm$0.07 & 30.1$\pm$0.3  & 4.7$\pm$0.3 & 9.3$\pm$0.2 & $-$11  & 2.30$\pm$0.02 \\ 
\enddata
\tablecomments{The time-integrated spectral analysis for Burst 1 was performed both over the whole burst (i) and for the second, brighter emission (ii). All statistical uncertainties are at the 1$\sigma$ confidence level. The t$_{start}$ and t$_{stop}$ are relative to the trigger time (in UTC). The fluence values are over \teq{10-1000}~keV using a COMPT spectrum. 
}
\end{deluxetable}\label{tab:1}

Comparing both burst spectra, the burst spectrum of Burst 1 is found to have a clear ingress and egress in \ep~overall (i.e., the envelope), peaking at around 50~ms after the onset of the second emission episode (25~ms in the right panel of Fig.~\ref{fig:1})\footnote{This soft-hard-soft evolution is suggestive of a radiation cone passing into and out of the line-of-sight of \fermi.}. This is markedly different from the flatter overall \ep~behavior of Burst 2 and likely implies different viewing geometries of the source environment. While a BB+BB model was not found to be the preferred fit to either burst spectrum, we calculate the BB region sizes to try to put constraints on the flux tube geometry. For Burst 1, we find the BB regions to be 283 $\pm$ 52 km$^{2}$ and 31 $\pm$ 3 km$^{2}$ for $kT_{1}$ and $kT_{2}$, respectively. Similarly, we calculate the BB regions for Burst 2 to be 359 $\pm$ 125 km$^{2}$ and 57 $\pm$ 8 km$^{2}$ for $kT_{1}$ and $kT_{2}$, respectively. The $kT_1$ region size is used to provide a constraint on the diameter of the cross section of the activated flux tube, as described in Section~\ref{sec:discuss_Ep}.

The light curve and \ep~time traces in Figure~\ref{fig:1} display similar character in their general shapes, or envelopes, suggesting that there is a significant correlation between the flux $\cal{F}$ and \ep. To explore this, we binned flux and \ep~data in 4~ms time intervals spanning the burst from the rapid risetime and decay of each event. A correlation of ${\cal F} \, \propto \, E_{\rm p}^{3.2 \pm 0.2}$ was found, with a Spearman rank order correlation coefficient and chance probability of $0.91$ and $5.50 \times 10^{-29}$, respectively.  This differed considerably from the ${\cal F} \, \propto \, E_{\rm p}^2$ correlation obtained by \citep{Roberts2021} for the GRB 200415A magnetar giant flare from the Sculptor galaxy, a correlation that is nominally a signature of Doppler beaming/boosting from ultra-relativistic outflows from a rotating star. Thus, we conclude here that the plasma that generated these two special bursts was likely moving only mildly-relativistically along the active field lines, with the radiation being only somewhat anisotropic at each emission locale. 

\newpage

\subsection{Periodicity Search for Burst 1}

We performed three complementary searches of the time series of \ep~values of Burst 1 at both $2~\mathrm{ms}$ and $4~\mathrm{ms}$ resolution for statistical evidence of a periodic or quasi-periodic oscillation (QPO); two Fourier approaches and a more sophisticated Gaussian Processes technique. As Burst 2 was found to be weaker than Burst 1, we did not search its time series for any spectral QPOs (abbreviated hereafter QPSOs\footnote{In order to differentiate between traditional QPOs and the spectral QPOs found in this study, we named the latter with the abbreviation, QPSO (quasi-periodic spectral oscillations) throughout the remainder of the paper.}) in this event.   
To search for periodic signals in the \ep\ lightcurves, we follow the procedure of \citet{huppenkothen2013}. We first generate a periodogram from the \ep~time series, fit a power law period distribution to this periodogram, and subsequently find the highest maximum outlier. The significance is calibrated using $1000$ simulated periodograms based on posterior distributions of the power-law model parameters generated using Markov Chain Monte Carlo simulations. We assumed wide, flat priors for the power-law index and the logarithm of the power-law amplitude, and a normal prior with a mean of $2$ and a width of $0.5$ around a noise value of $2$. We find no statistical evidence to reject the null hypothesis, namely that the periodogram is consistent with stochastic variability following a power-law power spectrum, and thus find no evidence for a periodic signal in the data using this analysis. 

Next, we search for QPSOs by comparing the power-law model for the periodogram to a model consisting of a power-law model and a QPSO -- parametrized as a Lorentzian -- through a Likelihood Ratio Test (LRT). Once again, the LRT is calibrated via $1000$ simulations generated from an MCMC sample of the null hypothesis (the power-law model). Here, we find weak evidence in the $2\,\mathrm{ms}$ data ($p=0.039$) that the observations might not be drawn from the null hypothesis. As the \ep~values are spectral model parameters associated with uncertainties, and because Fourier periodograms by default do not take uncertainties into account, we check the effects of incorporating the uncertainties through simulations. Here, for each point in the time series, we draw from a normal distribution with the mean given by the best-fit \ep~value, and the standard deviation given by the $1\sigma$ parameter uncertainties. In this way, we draw $1000$ time series. Subsequently, we compute the LRT between the models with and without a QPSO for each simulated time series, and compare the resulting distribution to the distribution generated from the MCMC simulations of only the power-law model. We find that, overall, while not exactly the same, the LRT of simulations generated from the best-fit values and their uncertainties have a similar distribution to the LRTs generated from the power-law model only, with no QPSO present. The periodogram of the best-fit values is a mild outlier ($p\sim 0.04$) with respect to both distributions, and thus we conclude that there is weak evidence for a rejection of the null hypothesis that the periodogram was generated by pure noise.

\begin{figure}
    \centering
    \includegraphics[width=\textwidth]{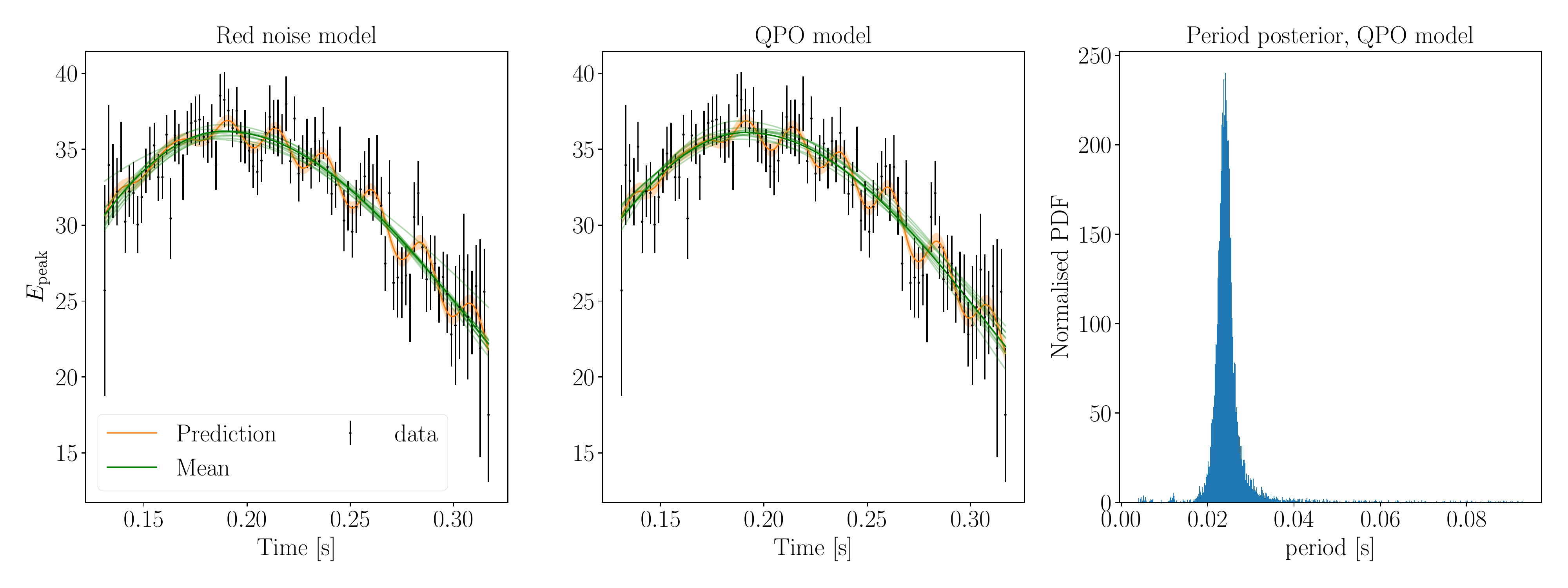}
    \caption{Searching for (quasi-)periodicities in \ep~with a Gaussian Process-based model. \textbf{Left:} E$_{p}$ values with uncertainties (black) as a function of time, along with posterior draws of the skew-Gaussian mean function (green) and posterior draws of the combined model of skew-Gaussian and damped random walk in orange. Overall, this model, containing only aperiodic stochastic variability on top of the overall skew-Gaussian trend, fits the \ep~very well. \textbf{Middle:} as in the left panel, but for a model where the damped random walk was replaced by a QPSO, parametrized as a stochastically driven, damped harmonic oscillator. This model, too, describes the data adequately well. The resulting Bayes factor does not allow us to favour one model over the other. \textbf{Right:} the posterior probability density for the period of the QPSO in the model shown in the middle panel, showing that the time series constrains the QPSO period well.} 
    \label{fig:2}
\end{figure}

Fourier-based methods have a number of shortcomings in this context. Firstly, they assume a stationary process, which the \ep~time series of a magnetar burst is not, and which is known to bias the significance of QPO detections \citep{huebner2021}. Secondly, as explained above, they cannot robustly take into account the uncertainties on the \ep~value. Finally, null hypothesis tests can only ever reject a null hypothesis, and thus cannot yield direct evidence \textit{for} the detection of a QPSO. With this in mind, we implement the Gaussian Process-based method recently introduced in \citet{huebner2022}. Gaussian Processes encompass a class of time domain-based models that addresses many of the above shortcomings. They enable direct (Bayesian) model comparison via comparison of model evidences in the form of Bayes factors. They can natively include non-stationarity via mean functions, and they include prescriptions to take into account data uncertainties. We choose a skew-Gaussian function as a mean model to account for the broad trend in the time series, and implement three different hypotheses for the variability seen on shorter timescales than that trend: a damped random walk in the \ep~variate to parametrize stochastic, aperiodic variability; a stochastically driven, damped harmonic oscillator as a QPSO model (corresponding to a Lorentzian in the Fourier domain); and a model combining both damped random walk and Lorentzian. 

For all models, we implement wide, uninformative priors as described in \citet{huebner2022}, sample the posterior of each model via Dynamic Nested Sampling \citep{dynesty}, and finally compute Bayes factors to compare all models. Unlike null hypothesis tests, Bayes factors directly compare two models to each other, and thus can be used to yield evidence for or against a given model. We find that the QPSO-only and the red noise-only models produce very similar evidences, $\log(B) = 0.018$ for both $2~\mathrm{ms}$ and $4~\mathrm{ms}$ time series. This is smaller than the uncertainties in the calculation of the evidences $\sigma_B \sim 0.2$, and both models thus describe the data equally well. We note that both are mildly preferred over the model containing both red noise and a QPSO, $\log(B) = 2$, likely owing to the increase in model parameters (and thus prior volume) of the more complex model. In Figure \ref{fig:2}, we present the $2~\mathrm{ms}$ time series along with posterior draws from the model containing only the mean function and red noise, and for the model containing only the mean function and a QPSO. We find that indeed, both are able to describe the observed data well. We also note, however, that the posterior probability density for the QPSO's period parameter is narrow and unimodal, $P = 0.02415^{+0.0023}_{-0.0018}$ seconds, suggesting that the QPSO is well-constrained. 

\section{Discussion}
 \label{sec:discuss}

To recap, the time-integrated spectra of the two selected bright bursts from \sgrb\ were found to prefer a non-thermal fit, and values of $E_{\rm iso}$ and $L_{\rm iso}$ for these two transients were comparable to those of typical energetic magnetar bursts. Our time-resolved analysis of \ep~in the brighter event (Burst 1) uncovered suggestive evidence of \ep~fluctuations, contrasting its lightcurve morphology which doesn't exhibit traditional QPO signatures. Careful and thorough analysis to evaluate the periodicity in these \ep~oscillations uncovered a well-constrained, 24~ms (i.e., 42~Hz) QPSO, the first time such possible periodic spectral fluctuations have been observed. Here we address some nuances of the QPSO analysis, summarize prior evidence for light curve QPOs in bursts, and deliver a theoretical interpretation for what could generate QPSOs in these two events.

\subsection{Bayes Factors in QPSO Probes}
 \label{sec:Bayes}

It is important to mention that there are several caveats concerning the Bayes factors adopted in the \ep~oscillation analysis.
First, the models that were compared were purely empirical choices, and in the absence of a physical model capable of generating the data we see, were our only option for searching for QPSOs in this kind of data. The damped random walk model makes strong assumptions about the data-generating process; it is worth bearing in mind that its standard assumptions originate in studies of black holes, not magnetars. We expect that a more physically-motivated model for generating the variability would render the Bayes factor a more reliable metric for model selection. Second, Bayes factors are known to be very sensitive to prior choices. Here we set wide, uninformative priors on the model parameters, so that our lack of physical knowledge of the data-generating process likely impacts the significance of the result. Third, a statistically sound analysis should rely on more than a single number. Here, we jointly considered the evidence from both the Fourier-based analysis, which yielded a mild outlier, and the Gaussian Process analysis, which yielded an inconclusive Bayes factor, but a large \textit{effect size} in the form of a narrow, well-constrained posterior on the QPSO period. For a classical noise process, we would expect the period posterior to largely reproduce the wide, flat, uninformative prior rather than be heavily concentrated and unimodal, as is seen here.

Finally, it is important to note the impact of chosen priors on the models to be compared, which in the Gaussian approach were the damped random walk and the Lorentzian QPSO. In the analysis above, we have considered the Bayes factor in isolation, which in practice is equivalent to assigning equal probability to both the damped random walk and the Lorentzian model. If QPSOs are physically expected, as discussed at length in Section~\ref{sec:discuss_Ep}, then priors change and the confidence in the existence of the 24 ms QPSO in Burst 1 rises accordingly. The combination of these considerations leads us to conclude that this candidate discovery is worth reporting, especially in the context of the observation and analysis of future bright bursts, where improved count statistics and longer durations may yield more confident detections.

\subsection{QPOs in Magnetar Bursts}
 \label{sec:magnetar_QPOs}

Quasi-periodic oscillations with frequencies of $\sim$18-625 Hz have been identified in light curves from Galactic magnetar giant flares from SGR 1806-20 and SGR 1900+14~\citep{Strohmayer2005,Strohmayer2006,Huppenkothen2014a,Miller2019}. Recently, QPOs have also been identified in the tail (7 to 160~ms) of extragalactic magnetar giant flare GRB 200415A \citep[183$\pm$20~Hz at 2.5$\sigma$ signficance, p-value of 2.3$\times$10$^{-2}\,$; see][]{Roberts2021}, and during the peak (3~ms) of GRB 200415A~\citep[2132 and 4250~Hz; see][]{CastroTirado2021}. \citet{Huppenkothen2014b} identified a weak QPO signal centered at $\sim$260~Hz in an SGR J1550-5418 burst, and more significant QPOs centered at $\sim$93 and $\sim$127~Hz when using stacking of the same data. A QPO centered at $\sim$57~Hz was also found by stacking 30 short, individual bursts from SGR 1806-20~\citep{Huppenkothen2014c}. More recently, a 40~Hz QPO at 3.4$\sigma$ (p-value of 2.9$\times$10$^{-4}$) was reported from \sgrb in data from the Hard X-ray Modulation Telescope (HXMT; an instrument on Insight) over an energy interval of 18-50 keV~\citep{Li2022}. This unusual burst was observed contemporaneously with FRB~200428 during the 2020 outburst, suggesting that some FRBs are related to strong oscillation processes in neutron stars.

All these QPOs were identified in the intensity of the lightcurve over frequencies predominately from 10–2000 Hz.  These are most likely associated with global torsional/axial oscillations of the neutron star, particularly in the lower frequency band below 150 Hz~\citep{Levin2007,Huppenkothen2014a,Miller2019}. However, models to explain spectral fluctuations (QPSOs) at similar frequencies are lacking, likely primarily because such a phenomenon has not been reported before.  Consequently, we now present an original theoretical interpretation of the QPSOs (\ep~oscillations) suggested by the spectro-temporal analysis above.

\subsection{Magnetar Burst Acoustics} 
 \label{sec:discuss_Ep} 

The emission region for the highly luminous bursts necessarily is highly optically thick to Thomson scattering due to the high densities of the radiating charges \citep[e.g.,][]{Lin2011b,Lin2012}. Thus the microphysical fluctuation (Thomson diffusion) timescales should be very small. In addition, the Alfv\'en mode frequencies lie below the ion plasma frequency $\omega_p = \sqrt{ 4 \pi e^2 n_p/m_p}$, which is likely $\sim 10^{14}\,$Hz for plasma number densities of $n_p \sim 10^{22}\,$cm${}^{-3}$ that can be deduced from burst radiation efficiency arguments \citep{Baring-2007-ApSS}. While very low frequency (long wavelength) Alfv\'en waves can be contemplated, the high plasma density effectively scatters and absorbs them. The result is a decoherence that indicates that Alfv\'en modes do not drive the \ep~oscillations that are clearly a macro-scale phenomenon. The high opacity implies that the emission environment should approximate, to a considerable extent, a quasi-thermodynamic ensemble of photons, pairs and possibly also ions.

The natural manifestation of thermodynamic fluctuations in such a highly magnetic environment is via density (\teq{\rho}) perturbations along surface-to-surface magnetic flux tubes accompanied by pressure (\teq{P}) variations mediated by the gaseous equation of state (EOS), i.e. acoustic oscillations. Pressure changes/fluctuations will then generate varying values for the observed photon \ep~value. Generally, the anticipated chaotic nature of the burst region should preclude clear oscillation signatures if the emission zone straddles many field lines. Yet, occasionally, the activated flux tube could possess a range of small cross sectional areas $A$, and then pressure wave signatures might not be totally obscured by the environmental chaos. The radiating plasma ``bounces'' between flux tube footpoints, with adiabatic expansion and compression along the tube seeding the \ep~oscillations. We explore here the implications of this picture.

A fairly ``constrained'' magnetic flux tube with a small cross section can serve as an acoustic cavity for the burst zone, so that natural oscillations can be permitted along its length on timescales of arclength \teq{\cal S} of the tube divided by the sound speed \teq{c_s}. This should be commensurate with the favored \teq{24}ms period in the QPO analysis of the \ep~variability. Observe, that the discussion above indicated that the historical explanations of QPOs in the light curves of magnetar giant flares \citep[e.g.,][]{Strohmayer2005,Strohmayer2006} have considered sub-surface seismic modes. In principle, these could provide a driver for the \ep~fluctuations. Yet, any periodic imprint of sub-surface drivers for burst activation is generally obscured by the high opacity of the flux tube, {\it unless} the seismic mode frequency approximately coincides with the natural frequency of the acoustic cavity, in which case the sub-surface driving is resonant.  This may or may not be the case for the bursts studied here.

Returning to the acoustics, since the radiation efficiency of the burst region is likely small, and the photon peak energies are well below the electron rest mass energy, the system pressure is putatively dominated by the gas contribution. The sound speed can then be {\it estimated} using the formalism provided by \cite{Synge-1957-book} for a 3D relativistic Maxwell-Boltzmann distribution of a single species. For the purposes of this discussion, we will presume that \teq{e^{\pm}} pairs constitute the gas. Note that the pair distributions strictly should be treated as one-dimensional due to rampant cyclotron/synchrotron cooling perpendicular to the strong magnetic fields. Yet addressing this nuance via the introduction of 1D Maxwellians does not qualitatively alter the conclusions drawn here.

For relativistic pairs, the sound speed is expressed in Eq.~(316) of \cite{Synge-1957-book} in terms of modified Bessel functions whose argument is \teq{1/\Theta}, where \teq{\Theta = kT/m_ec^2} is the dimensionless temperature. The observed values of \ep~ in the 30-35 keV range (see Figs.~\ref{fig:1} and~\ref{fig:2}) for the photons suggest \teq{kT\sim 10}keV for the temperature of the photon-pair conglomerate. This translates to \teq{\Theta \sim 0.02}, for which the sound speed approximately satisfies the familiar non-relativistic expectation \teq{c_s^2/c^2 \approx 5\Theta/3} (i.e. \teq{\gamma P/\rho} for adiabatic index \teq{\gamma = 5/3}). It then follows that the 3D sound speed realizes \teq{c_s\approx 0.176c}. The addition of moderate radiation pressure will increase this somewhat, but still fall well short of the relativistic EOS result \teq{c_s = c/\sqrt{3}},

With this evaluation, the natural timescale for sound propagation over a neutron star radius is \teq{\rns/c_s\sim 0.18}ms for pair plasma (with \teq{\rns = 10^6}cm). Accordingly, a \teq{24}ms fluctuation might signal acoustic propagation on a flux tube of length around \teq{{\cal S}\sim 130\rns}. The length might be somewhat smaller because of the adiabatic cooling along the flux tube that lowers both \teq{\Theta} and \teq{c_s} away from its footpoints at the stellar surface. Given magnetic flux conservation along such tubes, the cross sectional area $A$ rises at high altitudes and inversely as the field strength $B$, so that the plasma density scales roughly as $n_p \propto A^{-1}\propto B$ as it flows along a flux tube. The obvious deliverable here is that if a flux tube with a relatively small cross sectional area constitutes the active burst zone, then observations of \ep~fluctuations can provide a measure of the flux tube length. The arclength \teq{\cal S} of a dipolar field line from footpoint to footpoint in flat spacetime geometry is stated in Eq.~(23) of \cite{Wadiasingh-2018-ApJ}. This can easily be expressed in terms of the footpoint colatitude \teq{\theta_{\rm f}}, and in the domain of polar field lines, \teq{\theta_{\rm f}\ll 1} yields the approximation \teq{{\cal S} \approx 2.76 \rns/\theta_{\rm f}^2}. This dipolar result would indicate footpoint colatitudes \teq{\theta_{\rm f}\sim 8^{\circ}} for a \teq{24}ms fluctuation. This estimate increases somewhat (i.e. moving away from the pole) when twists are introduced to modify the field morphology, thereby lengthening field loops \citep[profoundly near the pole:][]{Hu-2022-ApJ}. This \teq{\theta_{\rm f}} estimate for an activated flux tube is broadly consistent with those obtained in axisymmetric plasma simulations of twisted magnetospheres \citep[e.g.,][]{Chen-2017-ApJ}. We note that if ions are abundant in the burst zone, then the sound speed will drop considerably, thereby decreasing the inferred flux tube length \teq{{\cal S}} and moving its locale somewhat towards the magnetic equator.  

The two-blackbody emission region area estimates of \teq{\sim 280-360}km$^2$ for the \teq{T_1\sim 5}keV cool Planck component (see Section~\ref{sec:oscill}) can be employed to estimate the typical cross sectional diameter \teq{d} of the flux tube; the cool region most likely corresponds to the major portion of the tube's length \teq{{\cal S}} centered on its equatorial apex. Assuming that the tube has an approximately circular cross section, its external observer-facing surface area is of order \teq{\pi {\cal S} \, d/2}, so that one deduces that the typical cross sectional diameter of the tube {\it near its apex} is around \teq{1.4-1.8\,\rns \ll {\cal S}}, so that the flux tube is appropriately slender. The corresponding tube diameter near its surface footpoints is a small fraction of \teq{\rns} for both dipolar and twisted field geometries.

Thus, with this intriguing and serendipitous observation, if \ep~fluctuations can be confirmed in bursts, they enable a path to discerning their emission locale. Yet the norm will be the circumstance where the active region possesses a large number of flux tubes of different lengths and spanning a larger cross sectional area \teq{A}, thereby providing destructive interference (noisiness) that obscures signature QPSO periods for acoustic oscillations. Thus, we expect that observations of \ep~oscillations will be rare in the magnetar burst database, as appears to be the case.

\section{Summary} 
 \label{sec:summary} 

In this study, we have presented the possible discovery of fluctuations of the spectral peak energy \ep~from two bursts from \sgrb during an outburst in early 2022. Time-integrated analysis of both bursts using \fermi~data found that their spectra were non-thermal. Time-resolved analysis of \ep~in both events provided initial evidence of \ep~oscillations (QPSOs) over the brightest part of each event, while their lightcurve morphology presented no indication of classical QPOs. Careful periodicity analysis supported this QPSO hypothesis through the suggestion of a well-constrained, 24~ms (42~Hz) QPSO in one of the events (Burst 1). We conjecture that these heretofore unprecedented events can be explained by density and pressure perturbations propagating along a highly magnetized flux tube that serves as an acoustic cavity. This tube is nominally of a length of \teq{{\cal S}\lesssim 130} neutron star radii for a purely pair plasma in the emission zone, though the \teq{\cal S} estimate drops considerably if ions are present and thereby lower the sound speed.

Both Burst 1 and Burst 2 occurred within \teq{\sim 4} days of each other, during the same outburst in January 2022. The spectroscopic and lightcurve data shown in Fig.~\ref{fig:1} show tantalizingly similar properties for the bursts. This suggests, perhaps speculatively, that both events and their acoustic oscillations arise either along the same flux tube, or a pair of flux tubes of comparable dimensions anchored on similar magnetic field lines. This would then provide an interesting constraint on the activation timescales over which the environment of the magnetosphere of \sgrb~(and perhaps, magnetars in general), change.

Detailed time-resolved analyses of other magnetar bursts are encouraged in order to find more examples of acoustic oscillations from magnetar sources. However, we note the difficulties in obtaining similar findings due to the need for magnetar bursts to be sufficiently bright and long enough in order to obtain enough oscillations or cycles, with which to confidently identify similar or varied QPSO frequencies above the noise level. This and the expected rarity of the proposed mechanism may make such instances rare when searching magnetar burst databases. Additionally, we use this study to advocate for physically accurate null hypothesis models to validate the detection of any future QPOs and QPSOs, as the current literature and models do not accurately describe magnetars sufficiently well.

\begin{acknowledgments}
O.J.R. gratefully acknowledges NASA funding through contract 80MSFC17M0022. 
M.G.B. acknowledges the generous support of the National Aeronautics and Space Administration through grants 80NSSC22K0777 and 80NSSC22K1576.
D.H. is supported by the Women In Science Excel (WISE) programme of the Netherlands Organisation for Scientific Research (NWO).
E.G. and Y.K. acknowledge the support from the Scientific and Technological Research Council of Turkey (T\"UB\.ITAK project number 121F266).
\end{acknowledgments}

\bibliography{sample631}{}
\bibliographystyle{aasjournal}


\end{document}